\documentclass[preprint2]{aastex}

\usepackage{txfonts}
\usepackage{natbib}
\usepackage{graphicx}
\usepackage{multirow}
\usepackage{amssymb}

\newcommand \cxc {{\it Chandra}}

\newcommand \bd {BD+30$^\circ$3639}
\newcommand \gvel {$\gamma ^2$~Velorum}
\newcommand \tmus {$\theta$~Muscae}
\newcommand{\kms}{km~s$^{-1}$}
\def\lsim{\mathrel{\rlap{\lower4pt\hbox{$\sim$}}
    \raise1pt\hbox{$<$}}}                % less than or approx. symbol
\def\gsim{\mathrel{\rlap{\lower4pt\hbox{$\sim$}}
    \raise1pt\hbox{$>$}}}                % greater than or approx. symbol

\shorttitle{Mixing of cold \& hot gas}
\shortauthors{Nordon et al.}

\begin{document}

\title{Narrow Radiative Recombination Continua: A Signature of Ions Crossing the Contact Discontinuity of Astrophysical Shocks}

\author{Raanan Nordon \altaffilmark{1,2}, Ehud Behar \altaffilmark{3,1}, Noam Soker \altaffilmark{1},} 
%\affil{Department of Physics, Technion, Haifa 32000, Israel} 
\email{nordon@mpe.mpg.de; behar@milkyway.gsfc.nasa.gov; soker@physics.technion.ac.il}
\and  
\author{Joel H. Kastner \altaffilmark{4}, \& Young Sam Yu \altaffilmark{4}}
\email{jhkpci@cis.rit.edu; yxy7181@cis.rit.edu}

\altaffiltext{1}{ Department of Physics, Technion, Haifa 32000, Israel}
\altaffiltext{2}{ Current address, Max-Planck-Institut f\"ur extraterrestrische Physik, postfach 1312, 85741 Garching, Germany}
\altaffiltext{3}{ Senior NPP Fellow, Code 662, NASA/Goddard Space Flight Center, Greenbelt, MD 20771}
\altaffiltext{4}{ Center for Imaging Science, Rochester Institute of Technology, Rochester, NY 14623-5604}

%\date{Received <date> Accepted <date>}

\begin{abstract} 
X-rays from planetary nebulae (PNs) are believed to originate from a shock driven into the fast stellar-wind ($v\sim$ 1000~\kms) as it collides with an earlier circumstellar slow wind ($v\sim$ 10~\kms).
In theory, the shocked fast wind (hot bubble) and the ambient cold nebula can remain separated by magnetic fields along a surface referred to as the contact discontinuity (CD) that inhibits diffusion and heat conduction.
The CD region is extremely difficult to probe directly owing to its small size and faint emission.
This has largely left the study of CDs, stellar-shocks, and the associated microphysics in the realm of theory. 
This paper presents spectroscopic evidence of ions from the hot bubble ($kT\approx 100$~eV) crossing the CD and penetrating the cold nebular gas ($kT \approx 1$~eV). 
Specifically, a narrow radiative recombination continuum (RRC) emission feature is identified in the high resolution X-ray spectrum of the PN \bd\
indicating bare C~VII ions recombine with cool electrons at $kT_e~= 1.7 \pm 1.3$~eV. 
An upper limit to the flux of the narrow RRC of H-like C~VI is obtained as well.
The RRCs are interpreted as due to C ions from the hot bubble of \bd\ crossing the CD into the cold nebula, where they ultimately recombine with its cool electrons.
The RRC flux ratio of C~VII to C~VI constrains the temperature jump across the CD to $\Delta kT > 80$~eV, providing for the first time direct evidence of the stark temperature disparity between the two sides of an astrophysical CD, and constraining the role of magnetic fields and heat conduction accordingly.
Two colliding-wind binaries are noted to have similar RRCs suggesting a temperature jump and CD crossing by ions may be a common feature of stellar-wind shocks.

\end{abstract}

%\maketitle

%%%%%%%%%%%%%%%%%%%%%%%%%%%%

\section{Introduction}

High velocity interactions among different astrophysical media due to collective plasma effects on length scales much smaller than the mean free paths for particle collisions are commonly referred to as collisionless shocks. For extensive reviews, see \citet{tidman71, mckee80}.
These shocks have been suggested to account for a wide variety of astrophysical phenomena ranging from the Earth's weak bow shock to strong shocks in supernova remnants (SNRs). Despite elaborate theoretical investigations, the detailed physics of collisionless shocks is far from being understood, mostly because direct observations or physical measurements of these shocks are notoriously difficult due to the extremely small size scales on which they occur.

The extended X-ray emission in planetary nebulae (PNs) is most commonly explained by a (reverse) shock driven into the fast wind expelled by the central star as it collides with the earlier slow and massive wind \citep[see review by ][]{frank99}.  High angular resolution X-ray images of PNs seem to support this notion as the X-ray gas referred to generally as the hot bubble appears confined to the interior of the optical and IR nebula \citep[e.g.,][]{kastner00}.
The low X-ray temperatures and observed luminosities cannot be explained by the present day wind velocities and mass loss rates. 
This perhaps suggests that the X-rays are due to the fast wind ejected during the early-PN phase or late post-asymptotic giant branch (post-AGB) phase, when the rapidly evolving wind was slower ($v~\approx $ 500~km~s$^{-1}$) and its mass loss rate was higher than what it is today \citep{akashi06}.  
Alternatively, conduction of heat from the fast wind to the slow wind has been invoked to explain the observed low X-ray temperatures \citep{soker94, steffen08}. 
In that case, it would be the nebular gas evaporating into the hot bubble that emits the X-rays.
The fast PN wind running into the denser and slower nebular gas leads to relatively slow shocks ($v_s~\approx $ 30~km~s$^{-1}$) propagating in this dense gas. 
In a few cases, the morphology of the PNs suggest that the fast wind is at least mildly collimated \citep{kastner03, sahai03}.

High resolution X-ray spectroscopy of PNs is particularly challenging owing to the low X-ray flux from these sources.
The only published X-ray grating observation of a PN is the \cxc /LETGS 300~ks exposure of \bd\ \citep{Yu2009,Yu2007}. 
The analysis of \citet{Yu2009} reveals a temperature range of (at least) 150--250~eV and extremely non-solar abundances. The elements C and Ne appear to be particularly enriched  with respect to solar abundances (C/O~$\sim$ 30, Ne/O~$\sim$ 4),
while Fe and N are deficient (Fe/O~$\sim$ 0.2, N/O~$\sim$ 0.4).
In fact, these abundances found in the X-ray plasma closely trace the abundances measured directly from the wind of the present-day WC central star of \bd\ indicating that the X-rays do indeed originate from the fast wind, while evaporation of nebular-composition gas and possibly also conduction of heat to the surrounding nebula have no observable effect \citep{Yu2009}.

The present paper is dedicated to an unusual emission feature in the LETGS spectrum of \bd\ around 25.30~\AA, which we interpret as a narrow radiative recombination continuum (RRC) of bare C~VII forming H-like C~VI. This feature, observable only at the high spectral resolution of the LETGS, provides a new insight into the region of the shocked fast wind of \bd\ and the microphysics at play. 

\section{Observations and Model}
\subsection{The Spectrum}
\label{Sec.Spec}

\begin{figure}[t]
\begin{center}
\includegraphics[width=1.0\columnwidth]{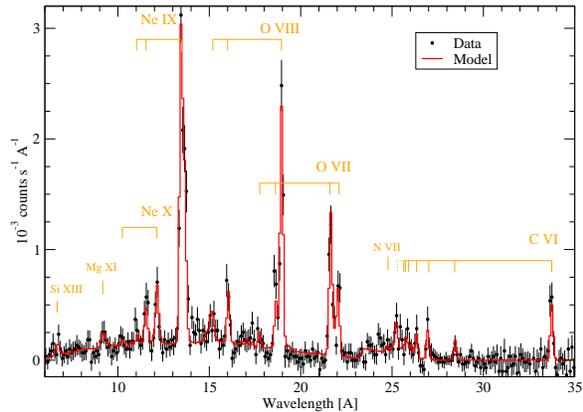} 
\caption{\label{fig:spectrum} LETGS background-subtracted spectrum of \bd, using bin size of 0.1~\AA, overlaid with the best-fitted model}
\end{center}
\end{figure}

\begin{table}[t]
\begin{small}
\caption{\label{tab:xspec_model} Parameters of the best-fitted model}
\bigskip
\begin{tabular}{|c|c|c|c|}
\hline
Parameter & Value $\pm$ Error & \citet{Yu2009}$^{\natural}$& Units \\
\hline \hline
N$_H$& 0.24 $\pm$ 0.04 & 0.24 $\pm$ 0.04 & 10$^{22}$ cm$^{-2}$  \\
\hline
\multicolumn{4}{|c|}{Thermal Components} \\
\hline
$kT_1$  & 260 $\pm$ 30  & 250 $\pm$ 30 & eV \\
$EM_1$ & 50 $\pm$ 10   & 70 $\pm$ 50 & 10$^{53}$ cm$^{-3}$ \\
$kT_2$  & 160 $\pm$ 30  & 150 $\pm$ 30 & eV \\
$EM_2$ & 160 $\pm$ 45 & 170 $\pm$ 35 & 10$^{53}$ cm$^{-3}$ \\
$A_C / A_O$ & 37 $\pm$ 8 & 33 $\pm$ 15 &  solar$^{\dagger}$ \\
$A_N / A_O $ & 0.3 $\pm$ 0.4 & 0.4 $\pm$ 0.4 & solar$^{\dagger}$ \\
$A_O $ & 0.9 $\pm$ 0.4 &  0.9 $\pm$ 0.4 & solar$^{\dagger}$ \\
$A_{Ne} / A_O$ & 3.8 $\pm$ 0.4  & 3.8 $\pm$ 0.8 & solar$^{\dagger}$ \\
$A_{Mg} / A_O$ &  0.8 $\pm$ 0.5 & 0.7 $\pm$ 0.5 & solar$^{\dagger}$ \\
$A_{Si} / A_O$ & 2.0 $\pm$ 1.1 & -	& solar$^{\dagger}$ \\
$A_{Fe} / A_O$  & 0.15 $\pm$ 0.1 & 0.2 $\pm$ 0.2 &  solar$^{\dagger}$ \\
\hline
\multicolumn{4}{|c|}{Recombination Component} \\
\hline
$kT_e^{cool}$ & 1.7 $\pm$ 1.3 & \nodata & eV \\
%$F^{RRC}_{CVII}$ & 8.8 $\pm$ 3 & \nodata & 10$^{-5}$ photons s$^{-1}$ cm$^{-2}$  \\
$F^{RRC}_{CVII}$ & 8.8 $\pm$ 3 & \nodata & \begin{tiny}10$^{-5}$ ph s$^{-1}$ cm$^{-2}$\end{tiny}  \\
\hline
%\multicolumn{4}{|l|}{$^{\natural}$ Upper and lower uncertainties have been averaged to facilitate comparison} \\
\multicolumn{4}{|l|}{$^{\natural}$ \begin{footnotesize}Upper and lower uncertainties have been averaged to\end{footnotesize}} \\
\multicolumn{4}{|l|}{\begin{footnotesize}$\quad$ facilitate comparison\end{footnotesize}}\\
%\multicolumn{4}{|l|}{$\dagger$ \citet{angr}} \\
\multicolumn{4}{|l|}{$\dagger$ \begin{footnotesize}\citet{angr}\end{footnotesize}} \\
%\multicolumn{4}{|l|}{{\it NOTE:} Abundance uncertainties are 90\% confidence range} \\
\multicolumn{4}{|l|}{\begin{footnotesize}{\it NOTE:} Abundance uncertainties are 90\% confidence range\end{footnotesize}} \\
\hline
\end{tabular} 
\end{small}
\end{table}

\bd\ was observed with \cxc\ for a total of $\sim$~300~ks, broken into several segments from February - December 2006. 
All observations were carried out with the low energy transmission grating spectrometer (LETGS) and advanced CCD imaging spectrometer (ACIS-S) configuration. The data were processed using the standard \cxc\ pipeline and combined into one spectrum using the standard tools of the CIAO software package. The full details of the observations and data reduction are given in \citet{Yu2009}.

As an initial step and following \citet{Yu2009}, we employed the XSPEC package \citep[version 12.3.1,][]{Arnaud1996} to fit the entire spectrum with two temperature components (2-T) of the astrophysical plasma emission code APEC \citep{apec}, photoelectrically absorbed by cold gas using the absorption  model of \citet{wabs}.
A minor improvement over the \citet{Yu2009} approach is introduced for modeling the spectral line profiles;
wavelength-dependent Gaussian smoothing is used.
Since \bd\ is $\sim4\arcsec$ across and spatially resolved by \cxc, and since LETGS is a dispersive slitless spectrometer, emission lines appear broadened in the spectrum due to the target extent.
This broadening is only slightly wider than the LETGS point-source line spread function and can be reasonably approximated by a Gaussian profile.
In order to test the Gaussian spatial broadening approximation, we fit for the energy dependence of the smoothing width $\Delta E \propto E^\alpha$.
The best-fit value obtained for $\alpha$ is  2.07 $\pm$ 0.15, consistent with spectral broadening of gratings in which $\Delta \lambda$ is roughly independent of $\lambda$, and thus $\Delta E \propto\Delta \lambda / \lambda ^2 \propto E^2$.

The best-fit spectral model is plotted over the source spectrum in Fig.~\ref{fig:spectrum}.
The most prominent lines in the spectrum are those of C, O and Ne K-shell ions. Fe L-shell lines, often the best temperature indicators, are in fact not clearly identified in the spectrum due to the low Fe abundance.
The best remaining temperature indicators are the line ratios of O VIII to O VII and of Ne X to Ne IX, which require here two non-degenerate temperature components of $kT=$160~eV and $kT=$ 260~eV.
The two temperatures likely represent a continuous temperature distribution in the hot bubble. For the case of a spherical wind, these range from the hottest inner region right behind the reverse shock, to the cooler, previously heated, more extended regions near the contact discontinuity (CD) that are still much hotter than the few-eV outer nebula \citep[see, e.g., Figs. 1 \& 2 in][]{akashi06}.
Much lower temperatures ($kT~<$ 100~eV) are usually difficult to constrain from X-ray spectra,
as the emissivities of most bright X-ray lines decrease strongly at these low temperatures.
This problem is augmented here by the significant photoelectric absorption ($N_H \cong 2.4\times 10^{21}~{\rm cm}^2$) toward \bd\ and by the decreasing LETGS throughput with decreasing energy.
The present spectrum, however, does provide unique temperature diagnostics as described in \S \ref{rrc} and in \S \ref{temp}.

Relative abundances can be accurately measured with X-ray emission line spectra. 
The measurement of absolute abundances $A_Z/A_H$ relative to hydrogen, on the other hand, 
is prone to systematic uncertainties as it requires assumptions for the H contribution to the bremsstrahlung continuum, which is harder to measure than narrow emission lines.
In the case of \bd, the low temperature (low bremsstrahlung intensity), the low H abundance, and the generally non-solar composition of the X-ray plasma resulting from core He-burning, all make the assessment of $A_H$ even more uncertain.
Consequently, we will limit our discussion in this paper to relative abundances,
which are presented in Table~\ref{tab:xspec_model}.
The present model parameters can be seen to very well agree with those of \citet{Yu2009}, as expected.
For more details on the uncertainties associated with the absolute and relative abundance determination, see \citet{Yu2009}.
For the purpose of the present analysis, it is sufficient to note the unambiguously high C abundance in the hot bubble of \bd.

\subsection{The Carbon Footprint of \bd}
\label{sec:Carbon_residuals}

\begin{figure}[t]
\includegraphics[width=0.95\columnwidth]{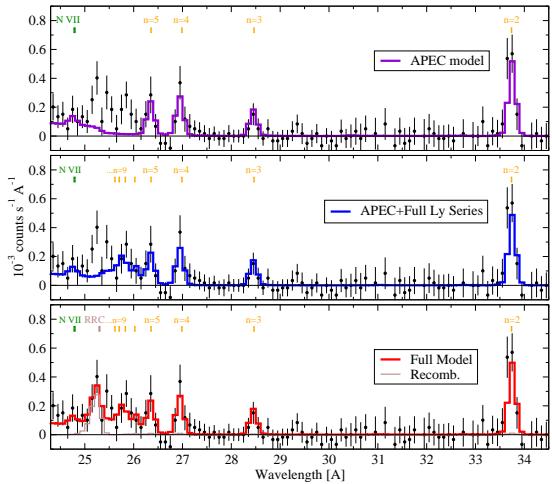} 
\caption{\label{fig:C_lines} Zoom into the carbon spectral region with 0.1~\AA\ bin resolution.
{\it Top:} Best-fit 2-T APEC model.
{\it Middle:} 2-T APEC model supplemented with the complete C~VI Lyman series. Note the failure of this model to fit the feature around the C~VII RRC at 25.30~\AA.
{\it Bottom:} The full model (red) including emission following recombination with cool electrons (thinner brown line). 
Note that interstellar absorption increases drastically with the wavelength across this range, which explains the skewed Lyman series line ratios. See Fig.~3 for the fluxed model.}
\label{fig_panels}
\end{figure}

The 2-T model described above fits the overall spectrum quite well, but fails to reproduce the significant emission of what appears to be a dense complex of spectral lines in the 25--26~\AA\ range.
These lines coincide in wavelength with the high-order C~VI Lyman series.
The top panel in Fig.~\ref{fig:C_lines} zooms in on the relevant spectral region showing the data and the best-fit 2-T model, while highlighting the C~VI Lyman series. 
The APEC model includes the C~VI Lyman series, but only radiative decays from levels with $n\leq 5$, $n$ being the principal quantum number of the upper level of the radiative transition.
%lines unaccounted for by the 2-T APEC model.
Thus, we first use the HULLAC code \citep{HULLAC} to complete the spectrum of the Lyman series up to $n=8$ explicitly, and the decreasing contributions of the remainder of the series by extrapolation.
The fitted model with the complete Lyman series is plotted in the middle panel of Fig.~\ref{fig:C_lines}.
Evidently, the agreement of the model with the observed data is somewhat improved.

Even after incorporating the complete Lyman series, the most prominent residual feature remains at 25.30~\AA.
We identify this feature as the narrow radiative recombination continuum (RRC) of bare C~VII recombining to form H-like C~VI.
In the following, we consider alternative identifications, but rule all of them out. First, this feature need not be confused with the N~VII Ly\,$\alpha$ line, which is very weak, but unambiguously resolved at 24.78~\AA\ (see Fig.~\ref{fig_panels}).
Indeed, \citet{murashima06}, using a lower resolution spectrum of \bd, reported a high N/O abundance ratio.
With the gratings, the two features are clearly distinct and the actual N/O ratio is much lower \citep{Yu2009}. 
Note that an overestimation of the N abundance should be expected from any low resolution observations that employ a model missing high-order C~VI lines.
Moreover, if a narrow RRC is present in the spectrum, it too would be confused at low resolution with N~VII Ly\,$\alpha$ and would lead to a further overestimate of the N abundance.

Other potential candidate lines around 25.30~\AA\ include the emission lines of L-shell ions of mid-Z elements such as Si, S, Ar, and Ca.  The L-shell ions of S have no bright emission lines between 25~-- 26~$\AA$ \citep{lepson05} and neither does Si (Lepson et~al. in preparation).
The strongest line of Li-like Ar~XVI is at 25.02~$\AA$ \citep{lepson03}, 
which given the LETGS resolving power,
is easily distinguished from the observed emission feature at 25.30~\AA. 
Moreover, Ar~XVI has equally bright lines at 24.87 and 23.53~\AA, neither of which is observed in the present spectrum. No other Ar ion has bright lines between 25~-- 26~\AA. Our atomic computations show that Ne-like Ca~XI has a 2p-4d line at 25.38~\AA. However, a Ca~XI 2p-3d line at least five times as strong is predicted at 30.47~\AA, and should be accompanied by the bright 2p-3s lines at 35.65, 35.67, and 35.79~\AA, none of which are observed in the spectrum.
Given the extremely non-solar abundances observed in some PNs, we also checked the spectrum of Sc. O-like Sc~XIV turns out to have a relatively bright 2p-3s line at 25.39~\AA.
However again, stronger lines of Sc~XIV that are not observed are expected based on our computations (e.g., a 2p-3d line at 22.95~\AA), as are lines of other L-shell Sc charge states. 

Charge exchange (CE) is another process that typically enhances high order lines when highly charged ions mix with neutral atoms. In the context of astrophysical shocks, unshocked neutral atoms can easily penetrate the CD (electro)magnetic barrier and mix with the hot gas.
If the enhancement is in sufficiently highly excited levels of C~VI ($n > 10$), the resulting lines could hypothetically fall below 25.5~\AA\ and would not be resolved from the C~VII RRC in the LETGS spectrum.
However, there are two strong arguments against such CE taking place in \bd. 
First, the UV radiation from the central star maintains high ionization (e.g., O~III) in the inner parts of the cool nebular gas just outside the CD.
Both \bd\ and its sister PN NGC~40 feature stratification in ionization, 
in which the innermost regions of the optical nebula, just outside the hot bubble, are the most highly ionized 
\citep{bryce99, sabbadin00}.
Observations (and an estimate of the ionization parameter $U \approx 0.03$) imply that the fraction of neutral species available for CE with the X-ray ions is less than $1/1000$.
The second argument against CE in \bd\ is that in order to produce an enhanced spectral signature below 25.5~\AA, the ionization potential of the neutral atom would need to be ridiculously low \citep[$\lesssim$ 2~eV, see][]{janev85}. 
This immediately rules out all of the abundant elements in \bd\ including He, C, O, and Ne (as well as H) by a large margin and essentially all other elements as well.

An intriguing possibility is CE with dust grains.
Dust grains can survive UV charging and may have low ionization energies (i.e., work functions), which could be of the order of a few eV.
Indeed, \citet{matsumoto08} found evidence of silicate grains in \bd\ on scales of $\sim 4\arcsec$ that coincide with the diameter of the X-ray determined CD.
However, at the corresponding distance of $4\times 10^{16}$~cm from the central star,
the grains inevitably are positively charged by more than a few elementary charges ($e$) as can be inferred from the work of \citet{feuerbacher73}.
This completely reduces the cross section for CE with positive (C~VII) ions. 
In short, we deem the CE process as highly unlikely to produce the emission around 25.30~\AA. 
All of this leads us to the conclusion that the 25.30~\AA\ emission feature must be the RRC of C~VII recombining to form C~VI. 

\subsection{The Carbon RRC}
\label{rrc}

\begin{figure}[t]
\begin{center}
\includegraphics[width=0.95\columnwidth]{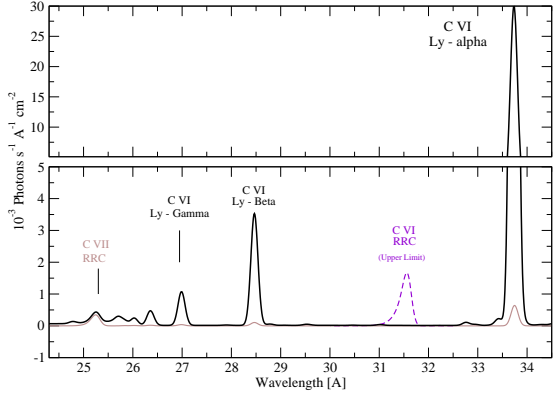} 
\caption{\label{fig:fluxed_model_spectrum} Fluxed best-fit spectral model for \bd.
The small contributions of C~VII recombination with cool electrons (as well as the upper limit for the C VI RRC - dashed line) compared with the strong line emission originating predominantly from hot gas demonstrate how hard it is to detect recombination from ions crossing CD surfaces in astrophysical shocks and how the extremely high C abundance of \bd\ ($A_C/A_O \approx 40$) is crucial for the present detection.}
\end{center}
\end{figure}

The narrow width of the RRC, which is a direct measure of the electron energy distribution in the plasma, indicates that recombination occurs with electrons of a few eV, two orders of magnitude colder than those required to ionize carbon to its C~VII state. 
In order to measure the exact recombining-electron temperature, we used the XSPEC {\it redge} model, which employs the flux density profile 

\begin{equation}
f_E = F^{RRC} \frac{1}{kT_e^{cool}} e^{-{\frac{h\nu - E_I}{kT_e}}} \quad \textrm{(photons s$^{-1}$ cm$^{-2}$ eV$^{-1}$)}
\end{equation}

\noindent where $F^{RRC}$ is the total photon flux in the RRC,
$kT_e^{cool}$ corresponds to the temperature in eV of recombining electrons,
$h\nu ~(= E_I +$ electron energy) is the photon energy, and
$E_I$ is the ionization potential of the recombined ion. 
The ionization energy of C VI was set at $E_I = 490.02$~eV and the two remaining independent parameters were fitted, yielding $kT_e^{cool} = 1.7 \pm 1.3$~eV and a total RRC flux of $F^{RRC}_{CVII}$ = (8.8 $\pm\ 3) \times 10^{-5}$ photons s$^{-1}$cm$^{-2}$.
The upper limit on $kT_e^{cool}$ is robust as the maximum width of the RRC is tightly constrained by the data.
Conversely, the lower limit is not as well determined due to the RRC being just broader than the instrumental line spread function expected from the extended angular size of \bd. 
At the distance of \bd, $d$=1.2~kpc \citep{Li2002},
we find the isotropic rate of C~VII recombination events resulting in photons of $\approx 25.30$~\AA\ (i.e., with cool electrons) to be

\begin{equation}
I^{rec}_{CVII} = 4\pi d^2 F^{RRC}_{CVII} = (1.5 \pm 0.5)\times 10^{40} \quad \textrm{s$^{-1}$}.
\label{RRCrate}
\end{equation}

For physical consistency, we also include in the model the C~VI line intensities due to recombination onto excited levels (up to $n=8$) and ensuing radiative cascades,
with relative contributions that are appropriate for $kT_e^{cool}$.
In the recombination model for $kT_e^{cool} = 1.7$~eV, the unabsorbed relative intensities of the RRC, Ly\,$\alpha$, Ly\,$\beta$, Ly\,$\gamma$, and Ly\,$\delta$ are 1.0, 1.35, 0.22, 0.08, and 0.04, respectively. 
These contributions were calculated using HULLAC \citep{HULLAC}, and here added to the model as narrow Gaussians at their fixed wavelengths.
The final model around the relevant spectral region can be seen in the lower panel of Fig.~\ref{fig:C_lines}, including the RRC and the low contribution of recombination to the lines compared to the high-$T$ emission.
Note that photoelectric absorption in this part of the spectrum is significant ($N_H = 2.4\times10^{21}$ cm$^{-2}$), and decreases strongly away from the C~I edge at $\sim 43.5$~\AA\ toward shorter wavelengths,
so that low order lines of the series (Ly\,$\alpha$, Ly\,$\beta$) are much more absorbed than the high order lines. 
The best-fit flux model corrected for interstellar absorption is presented in Fig.~\ref{fig:fluxed_model_spectrum} and the model parameters are listed in Table~\ref{tab:xspec_model}.

The miniscule flux due to recombination of highly ionized atoms with cool electrons compared with the strong line emission originating from hot gas makes the detection of the former tremendously hard. 
It is obvious from Figs. \ref{fig:C_lines} and \ref{fig:fluxed_model_spectrum} that if not for the unusually high C abundance ($A_C/A_O \approx 40$) in \bd, this detection would not have been possible.
Nevertheless, we did look for more evidence of recombination in the spectrum.
The RRC of C~VI forming C~V falls at 31.63~\AA.
Due to the low count rate at these long wavelengths, only an upper limit can be obtained, which is: $3.7\times 10^{-4}$ photons s$^{-1}$ cm$^{-2}$ (90\% confidence).
This flux is high (not very constraining), particularly when compared with the C~VII RRC flux of (8.8 $\pm\ 3) \times 10^{-5}$ photons s$^{-1}$cm$^{-2}$.
The difficulty is due to the low effective area of LETGS at this wavelength ($\sim 2$~cm$^2$) and the strong absorption near the C~I edge.
Nonetheless, this upper limit still proves useful, as will be demonstrated in \S \ref{sec:discussion}.
The flux in the O~VII forbidden line at 22.01 \AA\ and the He$\beta$ line at 18.63~\AA\ can be seen in Fig.~\ref{fig:spectrum} to exceed the flux predicted by the hot plasma model, which constitutes further, though tentative, evidence for recombination as the upper levels of these lines are preferentially populated by recombination and radiative cascades.
The O~VIII RRC forming O~VII falls at 16.78~\AA, but only an upper limit (90\% confidence) to the flux of $7.6\times 10^{-6}$ photons s$^{-1}$ cm$^{-2}$ can be obtained.
Blending with the Fe~XVII line at that wavelength is possible, but not expected to be significant, since Fe is considerably underabundant.
Although unambiguous detection is possible here only for the highly overabundant C, the upper limits are all consistent with the coolest ($kT \approx$ 100~eV) species in the hot bubble recombining with $kT \approx$ 2~eV electrons in the nebular gas.

\begin{figure}[t]
\begin{center}
\includegraphics[width=0.95\columnwidth]{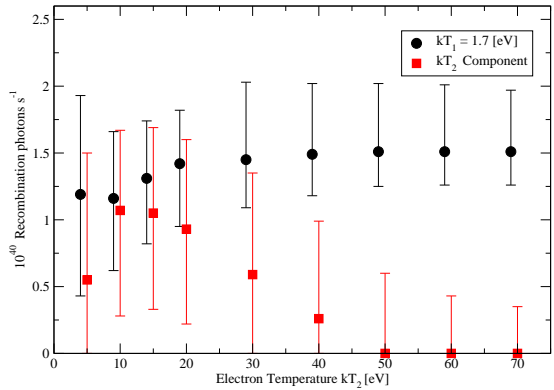}
\caption{\label{fig:Two_RRC}Constraints on C~VII recombination rates from two electron-temperature components as a function of $kT_2$ the temperature of the second component. The temperature of the first component is held constant at $kT=1.7$~eV.}
\end{center}
\end{figure}

The population of intermediate-temperature electrons ($kT \gsim 3$~eV) cannot be significant, as it would have smeared out the high contrast of the C~VII RRC.
To demonstrate this, we attempted to fit the same model as before, but with two C~VII RRC components instead of one.
We note that the quality of the data does not warrant a second RRC (see Fig.~\ref{fig:C_lines}).
Nevertheless, the main component was fixed at its previous best-fit value of $kT_1=1.7$~eV, while another cool component $kT_2$ was tested for (not to be confused with the high temperatures of the 2-T plasma model of \S \ref{Sec.Spec}).
The flux of both components was allowed to vary in the fit, as were the other model parameters.
In Fig.~\ref{fig:Two_RRC}, we plot 1$\sigma$ flux limits of each of the two RRCs as a function of $kT_2$. 
The 1.7~eV component is significant at all $kT_2$ values, corroborating the existence of cool electrons.
At high temperatures ($kT_2 > 20$~eV), the solution gradually converges to the single RRC result.
A 10 -- 20~eV RRC is possible, although it is consistently weaker than the 1.7~eV RRC (Fig.~\ref{fig:Two_RRC}).
Below $\sim$10~eV, the plot shows that there is a high degree of confusion between the two components. 
If we allow both RRC temperatures to vary, a best fit is obtained for $kT_1 = 1$~eV, $kT_2 = 10$~eV, $I_1 = 1^{+0.6}_{-0.5}$, $I_2 = 1^{+0.8}_{-0.7}$ $10^{40}$~recombination-photons~s$^{-1}$. These do not represent distinct components as the temperatures are highly confused and interchangeable. Only an upper limit of $kT<20$~eV can be obtained for the hotter components.
The addition of the second RRC (two extra free parameters) lowers the reduced-$\chi^2$ from 0.782 to 0.775 ($\Delta \chi^2 = 1.97$), if the method of \citet{Gehrels1986}  is used.
We conclude that recombination with electrons hotter than $\sim$~3~eV is not required by the data (see Fig.~\ref{fig:C_lines}).
However, insignificant recombination with electrons of up to $\sim$~20~eV cannot be ruled out.
A good probe of these intermediate temperatures could be O~VI emission \citep{gruendl04}, but
we are not aware of O~VI measurements for \bd.

%%%%%%%%%%%%%%%%%%%%%
\section{Discussion}
\label{sec:discussion}
%%%%%%%%%%%%%%%%%%%%%
In order to produce the observed narrow RRCs, highly charged ions have to collide with cool electrons at temperatures corresponding to only a few eV.
Narrow RRCs are typical of photoionized X-ray plasmas, where the ionization level is high, but the electron temperature remains low; good examples are X-ray binaries, or Seyfert~2 ionization cones.
In contrast to such sources, PNs have no hard enough X-ray source to ionize carbon up to the C~VII state. 
One might think that plasma instability around shocks could entail rapid radiative cooling in which ionized atoms recombine with cool electrons.
However, recombination is one of the primary cooling mechanisms, and highly charged species such as C~VII would have long recombined by the time the plasma cooled from above 100~eV to a few eV \citep[][Fig.~16 therein]{Sutherland93}, which rules out instability and rapid cooling as the origin of the RRCs.
In principle, adiabatic expansion could produce RRCs if it is sufficiently rapid that ions do not have time to recombine before the plasma cools to a few eV, that is, "ionization freezing". In reality, this is unlikely. The expansion time depends on the geometry, but the confined geometry of the hot bubble by the dense nebula does not favor such rapid expansion.
During the expansion, due to the adiabatic relation $T\propto n^{2/3}$, and since recombination rates  (Eq. \ref{trec}) scale as $nT^{-1/2} \propto T$, recombination would preferentially occur during the hot, dense phases and not during the low-density phases when the plasma cools down to a few eV.
Hence, it would not produce the observed narrow RRCs.
Moreover, for the hot gas to cool from 100~eV to 1~eV, its density would have to decrease by a factor of 1000.  Consequently, in order to produce the observed emission measure of the RRC ($n_e^2V \sim 10^{53} {\rm cm}^{-3}$), the size of the source would have to be more than two orders of magnitude larger than the hot bubble, which can be ruled out. 
A transient ionizing scenario, as expected for shocked gas, cannot produce RRCs either, although the initial ion temperature would be much higher than the electron temperature.
This is because electron impact ionization {\it follows} electron heating.
Even if one invokes a yet-unspecified mechanism to ionize the hot atoms (up to C~VII) before the electrons heat up, the collisions of cold electrons with hot ions much more efficiently heat the electrons than result in recombination.
We conclude that the most plausible physical origin of the observed RRCs is highly charged species that are heated and ionized in the hot bubble ($kT \gtrsim$ 100~eV) and then interact with much cooler electrons ($kT \sim$ a few eV) in a distinct medium. 

\subsection{Ions Crossing The Contact Discontinuity}
The most obvious reservoir of cool electrons lies in the cold nebula, where typical temperatures indeed correspond to a few eV.
The nebula is photoionized by the central UV source, so unbound electrons are abundant.
The observed recombination can take place, therefore, if ions from the hot bubble 
cross the CD plane and penetrate the slow wind where the cool electrons reside.
In the simplified hot bubble picture, the outer part of the hot bubble is also its densest and relatively coolest region \citep[for the spherically symmetric case see, e.g.,][]{akashi06, akashi07}.
This picture is consistent with the lowest charge states in the hot bubble crossing the CD and recombining, as the spectrum seems to suggest.
The narrow RRC and the consequential limits on intermediate temperatures should help constrain the role of heat conduction by electrons as well as the role of nebular evaporation into the hot bubble.

Heat conduction by electrons can be suppressed even by magnetic fields as weak as 0.1$\mu$G 
\citep{soker94}.
The viability of such magnetic fields has not been established, but if they do exist, they could preserve the steep temperature gradient across the CD required to produce a narrow RRC.
The length scales for such a gradient are of the order of the electron Larmor radius, while the Larmor radius for ions can be much larger. For particles with mass $m$ and charge $q$ at a temperature $T$, $R_L \sim \sqrt{mT}/q$.
Thus, ions can effectively penetrate the cold nebula across a magnetically held CD, while the opposite effect of cold electrons penetrating the hot bubble would be negligible.
The Larmor radius of a bare C VII ion with $kT^{hot} \approx$ 100~eV gyrating along a typical interstellar magnetic field component  of $B= 1~\mu$G (parallel to the CD surface) is 

\begin{equation}
R_L = 6\times 10^8 \left( \frac{B}{1\mu G} \right)^{-1} \left( \frac{kT^{hot}}{100 \rm{eV}} \right)^{1/2}  \rm{cm}
\label{rl}
\end{equation}

\noindent The corresponding half-circle (cross-back) rotation time would be:

\begin{equation}
\tau_L = \frac{\pi c M_C}{6eB} = 650 \left( \frac{B}{1\mu G} \right)^{-1} \rm{s}
\label{taul}
\end{equation}
 
\noindent where $M_C$ is the atomic mass of carbon. For comparison, the slow-down time by Coulomb collisions of a C~VII ion with the ambient cold electrons can be estimated from Equation (5-29) of \citet{spitzer56} to be (independent of ion temperature) roughly:

\begin{equation}
%\tau_s \approx 25 \left( \frac{kT^{cool}}{1 \rm eV} \right)^{3/2} 
%\left( \frac{n^{cool}}{10^4\mathrm {cm}^{-3}} \right)^{-1}
%\left( \frac{m^{cool}}{m_p} \right)^{-1/2}
%\rm s
\tau_s \approx 1200 \left( \frac{kT^{cool}}{1 \rm eV} \right)^{3/2} 
\left( \frac{n^{cool}_e}{10^4\mathrm {cm}^{-3}} \right)^{-1}
\rm s
\label{taus}
\end{equation}

\noindent where $n^{cool}_{e}$ is the number density of cools-gas electrons responsible for slowing down the fast C ions.
For the temperature and density of Eq. (\ref{taus}), if the C ion is stopped by protons $\tau _s \approx 800$~s (Spitzer 1956, Equation 5-28).
Obviously, the strength of the magnetic field and to a lesser extent the density in Eqs. (\ref{taul}, \ref{taus}) are fairly uncertain (even to an order of magnitude), but the fact that $\tau _s$ is comparable to $\tau _L$ implies that ions crossing the CD have a good chance of being stopped by the cool plasma, rather than returning to the hot bubble.
Chances can be much higher, if the magnetic fields are weaker than 1$\mu$G, or absent altogether.
Note that in order for the hot C ions to reach the 1.7~eV gas, they need to cross any intermediate region (say $kT = 10$~eV) without being stopped.
This considerably limits the size of such a region through $\tau_s (T)$ and the ion velocity,
which for intermediate temperature $kT = 10$~eV yields $\sim 10^{9}$~cm.  

Compared to these short time scales, the recombination of C~VII with the cool nebular electrons can take several months:

\begin{equation}
\tau_{rec} = 
	\frac{1}{n_e\alpha^{RR}(T_e)} \approx 1.8\times 10^7 \left(\frac{kT_e^{cool}}{1\rm{eV}}\right)^{1/2} 
	\left(\frac{n_e}{10^4 \rm{cm}^{-3}} \right)^{-1} \rm{s}
\label{trec}
\end{equation}

\noindent where $\alpha^{RR} = 5.5\times 10^{-12} {\rm cm}^3{\rm s}^{-1} (kT_e^{cool}/1~{\rm eV})^{-1/2}$ is the radiative recombination rate coefficient.
These timescales suggest the following plausible scenario:
ions from the hot bubble ($kT^{\rm hot}\sim$100~eV) cross the CD into the cool nebula.
Many of them slow down before they can gyrate back, and thermalize by collisions with the nebular electrons, cooling down to a few eV.
After a much longer time, and perhaps after they diffuse further upstream, the ions eventually recombine with the cool nebular electrons.

\subsection {Temperature Diagnostics at the Contact Discontiuity}
\label{temp}
As seen above, the width of the RRC can provide a tight constraint on the electron temperature on the nebular side of the CD. 
It would also be useful to measure the temperature of the hot plasma close to the CD, but just on its hot side.
This temperature can be readily obtained from the limit on the recombination rate of C~VI.
In a steady state, the rate of recombination with cool electrons is balanced by the ion crossing rate.
Indeed, steady state is reached within the typical recombination times $\tau_{rec} <$ year (Eq.~\ref{trec}), which are much shorter than the PN age.
The measured recombination rates $I^{rec}$, therefore, reflect the corresponding ion densities $n_{ion}$ and thus the fractional abundances in the hot gas $f_{ion}(T^{hot}) \propto n_{ion}$.
The C~VII to C~VI ionic density ratio is particularly sensitive to the temperature around $kT \approx 100$~eV.
These densities and temperatures refer to the edge of the hot bubble, right at the CD surface.
Taking into account that C~VII ions crossing the CD will produce two photons, one in the C~VII RRC and subsequently one in the C~VI RRC, one can write $I^{rec}_{CVII} \propto f_{CVII}(T^{hot})$ and $I^{rec}_{CVI} \propto f_{CVI}(T^{hot})+f_{CVII}(T^{hot})$, or: 

\begin{equation}
\frac{f_{C\,VII}(T^{hot})}{f_{C\,VI}(T^{hot})} = \frac{I^{rec}_{C\,VII}}{I^{rec}_{C\,VI} - I^{rec}_{C\,VII}}
\label{fratio}
\end{equation}

\begin{figure}[t]
\includegraphics[width=0.95\columnwidth]{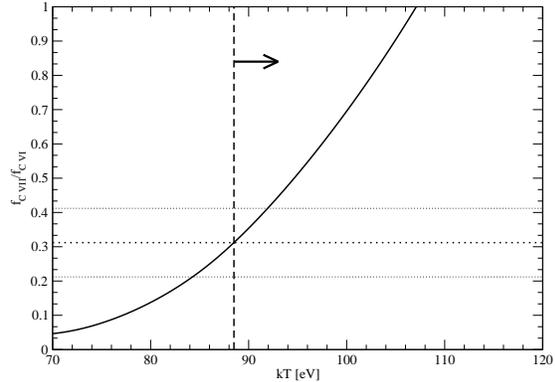} 
\caption{\label{fg:kT_from_fq}Ratio of C~VII and C~VI ionic fractions in collisional ionization balance, 
and as a function of the electron temperature (solid line).
The horizontal dotted lines indicate the lower limit for this ratio of $0.31\pm 0.1$ obtained from the C RRC flux ratio 
(Eq.~\ref{fratio}). 
The dashed vertical line marks the resulting lower limit on the hot bubble temperature in the immediate vicinity of the CD. }
\end{figure}

\noindent The measured upper limit for $I^{rec}_{C\,VI}$ (\S \ref{sec:Carbon_residuals}) places a lower limit on the ratio $f_{C\,VII}(T^{hot}) / f_{C\,VI}(T^{hot}) \geq 0.31 \pm 0.1$.
The theoretical ionic fraction ratio as a function of temperature \citep{Mazzotta1998} is plotted in Fig.~\ref{fg:kT_from_fq} along with the measured lower limit imposed by Eq.~(\ref{fratio}).
These results indicate that the temperature on the immediate hot side of the CD must be at least $kT^{hot} \geq 88 \pm 4$~eV.
Together with the accurate measurement of $kT^{cool}~= 1.7 \pm 1.3$~eV from the width of the RRC, this implies a temperature jump corresponding to {\it at least} 80~eV across the discontinuity, with no significant plasma at intermediate temperatures.
In principle, the O RRCs could have provided independent diagnostics, perhaps even a valuable upper limit on $T^{hot}$, since the $A_C/A_O$ abundance ratio in the hot bubble is well constrained.
Unfortunately, the present upper limit on the O~VIII recombination is not restrictive enough to provide meaningful constraints.

The requirement for a steep temperature jump corresponding to $\sim 80$~eV across the CD is robust and it raises an intriguing question regarding the role of magnetic fields and heat conduction. 
Heat conduction would drastically change the temperature and density profiles across the CD.
Instead of a sharp-contrast interface, a more gradual conduction front would manifest the transition between the hot bubble and the nebula as nebular gas evaporates into the hot bubble.
On the face of it, this picture seems to be in contrast with the LETGS spectrum of \bd\ and the present findings. 
In addition to the sharp temperature jump across the CD implied by the present analysis,
the fast-wind composition of the hot bubble in \bd\ argues against a significant role for nebular evaporation into the hot bubble, at least in this source \citep{Yu2009}.
Note, however, that the penetration of hot bubble material into the nebula will alter the chemical abundances of the immediate inner parts of the nebula, eventually changing the composition to that of the fast stellar-wind.
Such abundance gradients in the inner nebular regions, gradually changing from fast-wind composition to nebular composition as the distance from the center increases, were indeed found by \citet{sabbadin00} in spatially resolved observations of NGC~40. 

\citet{steffen08} published numerical simulations for PN hot bubbles that include heat conduction in the magnetic-free limit.
Strictly speaking, these models cannot be applied to \bd, since they assume solar abundances dominated by hydrogen, while the X-ray gas in \bd\ considerably departs from solar composition,
 consists predominantly of He and C, and is extremely H-deficient.
Nevertheless, we note that the temperature profiles of \citet{steffen08} at the outer edge of the conduction front do appear to sustain a sharp temperature drop by approximately two orders of magnitude (see their Fig. 4), perhaps due to the rapid cooling of plasma between a few eV and 100~eV.
However, the spatial resolution of these simulations is of the order of $10^{15}$~cm, while the thickness of the transition layer estimated in the present work is several times $10^{13}$~cm (\S \ref{sec:Further_Estimates}).
%, thus severely limiting the role of heat conduction.
Careful modeling of heat conduction on the small scales probed here and under conditions closer to those of \bd\ (He and C dominated plasma) will be needed before making a more conclusive and quantitative statement regarding the role of magnetic fields and heat conduction in PNs with WC central stars.

\subsection{Further Estimates: Carbon Mass Outflow, Densities, and Penetration Depth}
\label{sec:Further_Estimates}
The measured RRC intensity (Eq.~\ref{RRCrate}) implies a CD mass crossing rate of $\dot{M}_{C\, VII} =(4.7 \pm 1.6)\times 10^{-9}$ M$_\odot$~yr$^{-1}$,
and further dividing by the fractional abundance of C~VII at 100~eV \citep[$f_{CVII}$=0.27,][]{Mazzotta1998}, yields a total recombining C mass rate of $\dot{M}_C \approx (1.8\pm 0.6) \times 10^{-8}$ M$_\odot$~yr$^{-1}$. 
We recall that approximately 50\% of the hot bubble mass of \bd\ is likely in the form of C \citep{Yu2009, marcolino07}.
If this estimate holds near the CD, it implies a mass crossing rate of $\dot{M} \approx (3.5\pm 1.2) \times 10^{-8}$ M$_\odot$~yr$^{-1}$.
The most recent fast-wind mass outflow rate estimate for \bd\ is $5\times 10^{-7}$ M$_\odot$~yr$^{-1}$ \citep{marcolino07}, which is an order of magnitude less than the estimate of \citet{Leuenhagen1996}.
Therefore, the CD mass crossing rate is, at the most, less than 10\% of the mass outflow rate  
and even a smaller fraction of the early PN wind that had an even higher $\dot{M}$.
To that end, the ions crossing the CD seem not to have an appreciable effect on the stellar-wind, or the shock dynamics.

With a simplifying assumption of a smooth CD geometry that ignores wind clumping \citep[see][]{prinja07},
%As a consistency check
the ion density can be inferred from the observed recombination rate.
We assume that ions cross from the hot bubble across the CD into the cold nebula exclusively due to their thermal motion.
As in the previous section, 
%suming the physical conditions do not change significantly over the recombination time scale of a few months, 
in a steady state in which the C~VII crossing rate is balanced by the observed recombination rate $I^{rec}_{C\,VII}$, one can write: 

\begin{equation}
I^{rec}_{C\,VII} = 0.5 v_\perp S_{CD} n_{C\,VII}^{hot}
\label{ss}
\end{equation}

\noindent where $v_\perp = \sqrt{kT^{hot}/M_C}$ is the average magnitude of the C~VII velocity component perpendicular to the CD, whose surface area is $S_{CD}$.
The factor of 0.5 comes from the isotropy of the thermal motion. 
%This is the fastest possible crossing rate for the ions.
Eq. (\ref{ss}) can then be used to express the C~VII ionic density on the hot side of the CD as follows:

\begin{equation}
n_{C\,VII}^{hot} \approx 0.5 
%\left(\frac{I_{rec}}{1.5\times 10^{40} \rm{s}^{-1}} \right) 
	\left(\frac{S_{CD}}{2\times 10^{34}~\rm{cm}^2}\right)^{-1}
	\left(\frac{kT^{hot}}{100~\rm{eV}}\right)^{-1/2} 
	\rm{cm}^{-3}
\label{nion}
\end{equation}

\noindent where we assumed the CD surface area $S_{CD}$ to be a perfect sphere with the observed X-ray radius of \bd : $R_{CD}=4\times 10^{16}$~cm ($2\arcsec$ at 1.2~kpc). Using again the ionic fraction $f_{CVII}$(100 eV)~= 0.27 yields a C density of $n_C \approx 2$ cm$^{-3}$. 
Assuming as before and following \citet{marcolino07} and \citet{Yu2009}, a non-solar composition of gas depleted of hydrogen and dominated by He and C ($A_{C}/A_{He} \approx 0.4$ by number), we can estimate the electron density in the hot bubble to be $n_e \approx 3n_{He} \approx 8 n_C \approx 16$ ~cm$^{-3}$.
To that extent, the observed recombination rates appear to be consistent with the scenario of shocked hot bubble gas penetrating the slow wind, and with the theoretical models and numerical simulations of such a scenario.

Of course, the assumption of spherical symmetry in these models and in Eq.~(\ref{nion}) is probably an oversimplification. Collimated fast winds would imply higher densities.
Furthermore, the simplified picture that only includes crossing of hot ions to the cool side must be incomplete as it would breach the plasma neutrality and would very quickly induce a strong electric field that would inhibit further ion crossings.
Hence, there must be a reverse-charge current, for example of electrons following the ions, or nebular ions entering the hot bubble.
Further discussion of the complex morphology of the electric and magnetic dynamical structure around the CD is clearly beyond the scope of our study,
but these findings could be useful to constrain such investigations in the future.

Finally, the C~VII RRC intensity, along with the above estimate of the density, can be used to approximate the penetration depth $\Delta R$ of the hot bubble material into the nebula.
Using the fact that the recombination rate $I_{CVII}^{rec}$ scales with the electron and ion densities integrated over the entire emitting volume,
which in turn can be approximated by a thick sphere ($S_{CD}$) of width $\Delta R$, just outside the observed X-ray hot bubble (the CD), one can write

\begin{equation}
\Delta R = \frac{I_{CVII}^{rec}}{n_e \alpha^{RR} S_{CD} \left< n_{CVII} \right> ^{cool}} 
\end{equation}

\noindent where $\left< n_{CVII} \right> ^{cool}$ is the steady-state average C~VII density on the cold side of the CD.
This average likely represents a density gradient that 
cannot be constrained without a valid model for diffusion and plasma effects in the nebula.
All we can do at this point is scale $\Delta R$ with $\left< n_{CVII} \right> ^{cool}$, 
which can be arbitrarily equated with $n_{C\,VII}^{hot}$ of Eq.~\ref{nion}, to yield

\begin{equation}
\Delta R \approx 3 \times 10^{13}
	\left(\frac{n_e}{10^4~\rm{cm}^{-3}} \right)^{-1}
	\left(\frac{kT_e^{cool}}{1~{\rm eV}} \right)^{1/2}
	\left(\frac{kT^{hot}}{100~\rm{eV}}\right)^{1/2} 
{\rm cm} 
\end{equation}

\noindent At the distance of \bd, this is only $\sim$~1mas on the sky, which clearly cannot be resolved by X-ray telescopes even if the actual value of $\left< n_{CVII} \right> ^{cool}$ is much smaller.  
Such a shell needs to cool rapidly enough to radiate away the heat due to the penetration of the hot gas. 
A penetration rate of $\sim 10^{42}$ particles per second (c.f., Eq. \ref{RRCrate}), each contributing 100~eV, yields a heating rate of $\sim 10^{32}$~erg~s$^{-1}$.  
If the density in the nebula is $\sim 10^4$~cm$^{-3}$, a shell at a temperature of $kT \approx 2$~eV and with a volume of $S_{CD}\Delta R \approx 6\times 10^{47}$~cm$^3$ would cool at a rate of at least 6$\times 10^{33}$~erg~s$^{-1}$ \citep[given $\Lambda > 10^{-22}$~erg~cm$^3$s$^{-1}$,][]{Sutherland93}, safely above the heating rate.
Turning the argument around and requiring that the shell be able to expel the heat from penetrating ions constrain its thickness to a more conservative value of $\Delta R > 5\times 10^{11}$~cm.

There is reason to believe the stellar-wind material continues to diffuse into the nebula much after it recombines as spatially resolved images of NGC~40 show abundance gradients in low-ionization species on scales of a few arcseconds, or a few times $10^{16}$~cm. 
The penetration of the stellar-wind from the hot bubble deep into the nebula is a different effect from nebular evaporation,
although both processes similarly defy the simplified notion of stark disparity in composition between the two media.
In contrast with the chemical composition, however, the temperature disparity does appear to preserve the discontinuous nature of the interface.

\subsection{Ion Crossing of the Contact Discontinuity as a
General Phenomenon in Stellar-Wind Shocks}
The present discovery of a narrow carbon RRC in a PN shock is possible owing not only to the cold, yet ionized gas of the ambient nebula, but also to the extremely high C abundance in the stellar-wind and the hot bubble.
It would be natural to suppose, however, that hot ions crossing the CD should be a more general phenomenon of stellar-wind shocks. 
Indeed, we found two more cases of narrow RRCs in hot plasma sources:
both are colliding wind binaries with a C-rich Wolf-Rayet (WR) star and an O star companion: \gvel\ and \tmus.

\gvel\ is a WC8+O7.5 stellar binary system, which like \bd\ (WC9) is highly enriched in C, O, and Ne. 
\citet{schild04} reported the detection of C~VII and C~VI RRCs in the X-ray spectrum of \gvel,
and upper limits to the RRCs of oxygen ions.
Similar to \bd, the observed recombining gas in \gvel\ reflects the WR chemical composition.
In the binary case, the O star wind plays the role the cold nebula plays in PNs,
essentially stopping the WR wind and sending back a reverse shock.
The electron temperature of the C~VII RRC measured by \citet{schild04} is $3.3 \pm 0.7$~eV, 
consistent with that of an unshocked O star or WR wind.
The RRCs in \gvel\ persist when the thick WR wind absorbs the inner, hot X-ray components implying that they originate downstream from the stagnation point along the shock front \citep[e.g., region 3 in Fig.~9 of][]{schild04}.
\citet{schild04} speculated that the narrow RRCs could be due to cold plasma in extended regions of the wind, photoionized by hard X-rays from the hot wind-collision region.
However, they also noted significant problems in finding a model that can account for both the high ionization parameter required to photoionize C to its highest charge state (suggestive of low density) and the high emission measure of the recombining plasma (high density). 

\citet{schild04} also raised the possibility that recombination may occur following adiabatic cooling of the shocked gas.
As discussed in \S \ref{sec:discussion}, this would require the (adiabatic) dynamical times to be considerably shorter than the recombination times, which is highly unlikely in a dense stellar-wind.
Using the parameters of \citet{schild04}, we estimate the density 1~AU away from the WR star to be $n_e \approx 4\times 10^8 {\rm cm}^{-3}$, which implies C~VII recombination times of $\tau_{rec} \approx 5000$~s (for $kT = 100$~eV).
Adiabatic expansion times are harder to estimate without knowing the geometry, but the dynamical time it takes plasma that cools down to $kT = 1$~eV ($C_{sound} \approx 10$~\kms) to expand to say $\sim$1~AU is roughly 10$^7$~s $\gg \tau_{rec}$.
As in the PN case discussed in \S3, if the hot gas expands and cools from $kT > 100$~eV to a few eV, the density would have decreased by a factor of $\sim 1000$.
This would require an enormous volume to produce the observed RRCs.
We suggest alternatively that the RRCs in \gvel, in analogy with the case of \bd, are due to highly ionized C atoms that were shocked in the WR wind, crossed the CD surface, and interacted with unshocked electrons of the much cooler, yet also ionized plasma from the O star wind. The high carbon abundance of the WR wind, as opposed to the much lower abundance in the O~star wind, suggests that the hot ions must come from the WR side. The wind from the O star is also shocked (and hot) close to the stagnation point and therefore this hot-WR-ions/cold-O-electrons interface must exist farther away along the bow as \citet{schild04} inferred from observations.

\citet{sugawara08} identified narrow C~VII and O~VIII RRCs in the X-ray spectrum of another colliding-wind binary, \tmus, a WC6+O9.5 system fairly similar to \gvel. 
They suggested the possibility of ions escaping the bow shock layer and interacting with the wind, but provided no further discussion.
Again, we suggest the cool recombining gas interacting with the highly ionized plasma in \tmus\ be interpreted as evidence for shock heated plasma in the metal-rich WR wind crossing the CD and mixing with the cooler O star wind.  
The bright RRCs in the two stellar binary systems and in \bd\ suggest this kind of mixing and perhaps other microphysical processes around the CD are common to stellar-wind shocks in different systems ranging in size from $\sim 10^{13}$~cm in stellar binaries to $10^{16}-10^{17}$~cm in PNs. 
The high C abundance in these sources obviously facilitates the detection of the RRCs of C~VII and C~VI, which are relatively isolated in the spectrum. O RRCs are harder to identify unambiguously owing to blending with Fe L-shell lines.
We suspect, nonetheless, that the observed RRCs are traces of shocked plasma crossing the CD irrespective of element biases.

An interesting question is whether similar CD crossing takes place in young shell-like SNRs and whether it leaves a measurable signature that can give an idea of the temperature jump across the CD, as it does in \bd\ and in the colliding binary winds.
The fast stellar-wind colliding with the PN is qualitatively analogous to the SN ejecta slamming into the circumstellar material in young SNRs.
%Indeed, the temperature behind the reverse shock in young SNRs decreases towards the CD \citep[e.g.,][]{decourchelle01}. 
However, due to the high velocity of the SN explosion and the low density of the circumstellar gas,  temperatures on {\it both} sides of the CD, heated by the forward and reverse shocks, are both expected to be high ($kT~>$ 100~eV and possibly much higher).
Additionally, the immediate region around the CD represents the oldest shocked plasma, so the electrons that were initially cool have had time to equilibrate with the hotter protons. 
Consequently, ions crossing the CD would not produce a narrow RRC even if they recombine.
Moreover, unlike in PNs, the influence of strong magnetic fields on the CD in SNRs is not in doubt,
as indicated by radio synchrotron emission from thin filaments attributed to the CD region.
The interstellar magnetic field, and in particular its tangential component, can be enhanced by a factor of a few at the CD due to field compression by simple charge advection \citep{cassam05}, or by up to two orders of magnitude due to hydrodynamical instabilities \citep{jun95}. 
These fields are more than sufficient to sustain the temperature jump and perhaps even to totally suppress ion crossing.
By substituting a magnetic field of 100~$\mu$G and a density of 1~cm$^{-3}$ into Equations.
(\ref{taul}) and (\ref{taus}),
one easily sees that a thermal ion gyrating across the CD of a SNR will spend far too little time on the cool side of the CD ($\tau _L \sim $ seconds) in order for it to significantly cool down and remain on the other side ($\tau _s \sim $ days), even if cool electrons were present.
Indeed, the highest quality published X-ray grating spectra (especially beyond 25~\AA) of young SNRs do not appear to feature detectable narrow RRC emission \citep{andy01, behar01}.
We therefore conclude that narrow RRCs due to ion crossing of the CD are likely a unique signature of relatively slow shocks in dense stellar-wind sources.

%%%%%%%%%%%%%%%%%%
\section{Conclusions}
%%%%%%%%%%%%%%%%%%
\bd\  was observed with \cxc\ LETGS for a total of 300~ks providing the first and so far only high-resolution X-ray spectrum of a PN. 
As already shown by \citet{Yu2009}, the spectrum can be fitted by an absorbed, two-temperature plasma model, though it likely represents a distribution of temperatures between $\sim$100 -- 300 eV. 
The spectrum is dominated by C, Ne, and O emission lines with very little N and Fe. 
The carbon abundance is particularly high, indicating that the X-ray gas primarily originates from the present-day WC stellar-wind and making the C emission features most suitable for detailed analysis.

We detect in the spectrum a narrow RRC of bare C~VII forming C~VI by recombination with cool ($kT_e = 1.7 \pm 1.3$~eV) electrons.
We suspect that a bright RRC of H-like C~VI forming C~V also appears in the spectrum, but the low S/N of that feature allows only an upper limit to its flux.
We interpret the RRCs as direct evidence of penetration of hot, highly ionized plasma into the cool nebula past the CD, which separates the heated fast wind from the slow wind.
This finding requires a steep temperature gradient, as intermediate electron temperatures higher than $kT_e = 3$~eV would have broadened the RRC beyond its observed width. 
Using the measured lower limit to the flux ratio of the C~VII and C~VI RRCs, the temperature difference between the hot and cold plasma across the interface is found to be $\Delta kT > 80$~eV.
Such a steep gradient can be sustained by magnetic fields, which would significantly suppress heat conduction. 
On the other hand, magnetic fields may not be needed to preserve the sharp temperature drop between the hot bubble and the nebula. Numerical simulations of heat conduction in the literature \citep{steffen08} do appear to preserve a sharp gradient, though their grid is too coarse to resolve the processes discussed above.
Simulations with the appropriate WC-wind composition and at higher resolutions will need to be confronted with the present results in order to quantitatively test the role of magnetic fields in PNs. 

The measured recombination rate of $(1.5 \pm 0.5) \times 10^{40}$~s$^{-1}$ implies a mass crossing rate of $\sim3.5 \times 10^{-8}$~M$_\odot$~yr$^{-1}$, which can be a few percent of the total fast-wind mass,
implying that the fast-wind gas successfully makes its way through the nebula despite the putative magnetic barrier.  
The deep penetration of stellar-wind gas into the optical nebula is indirectly supported by abundance gradients observed in the innermost regions of NGC~40 \citep{sabbadin00}.
This picture is quite different from that of nebular evaporation into the hot bubble for which we find no observable sign.
Finally, we note two stellar (WC + O) binary X-ray sources with similar RRC features.
We take this as evidence of the generality of ions crossing CD surfaces in slow, dense stellar-wind shocks, regardless of whether the shock front is on the scales of a massive-star binary system or a PN.
The hotter and more highly magnetized environments of CDs in SNRs, on the other hand, are less favorable for ion crossing, or for producing narrow RRCs.

\begin{acknowledgements} 
We thank Ari Laor, Maurice Leutenegger and John Raymond for useful discussions and comments.
The research at Rochester Institute of Technology was supported by NASA through {\it Chandra} award GO5-6008X issued by the {\it Chandra} X-ray observatory center, which is operated by the Smithsonian Astrophysical Observatory for and on behalf of NASA under contract NAS8-03060.
\end{acknowledgements}

%%% BIBLIOGRAPHY %%%%%

%%% ONLINE APPENDIX %%%%%
%\Online
%\begin{appendix}

%\end{appendix}

\end{document}